\documentclass[iop]{emulateapj}		% For AJP manuscript journal format
\usepackage{epsfig}
\usepackage{epstopdf}			% For eps figures, old commands
\usepackage{graphicx,color}		% For eps figures, newer & more powerfull
\usepackage{amssymb}			% For useful mathematical symbols
\usepackage{color}			% For color text: \color command
\usepackage{url}			% For breaking URLs easily trough lines
\usepackage{amsmath}			% Provides various features to facilitate writing math formulas
\usepackage{rotating}			% For rotate any object of an arbitrary angle
\usepackage{float}			% For improveing the interface for defining floating objects
\usepackage{textcomp}			% Provides many text symbols (such as baht, bullet, copyright,etc )
\usepackage{psfig}
\usepackage{dcolumn}
\usepackage{times}
\usepackage{tabularx}
\usepackage[colorlinks=true,citecolor=blue]{hyperref}
\usepackage[english]{babel}
\shorttitle{Dynamics of small bright dots in the transition region above sunspot}
\shortauthors{T. Samanta et al.}

\begin{document}

%%%%%---------------Title----------------%%%%%
\title{Dynamics of subarcsecond bright dots in the transition region above sunspot and their relation to penumbral micro-jets}

\author{Tanmoy Samanta$^{1}$,
Hui Tian$^{2}$,
Dipankar Banerjee$^{1,3}$, 
% Yukio Katsukawa$^{4}$, 
Nicole Schanche$^{4}$}
\affil{$^{1}$Indian Institute of Astrophysics, Koramangala, Bangalore 560034, India; {\color{blue}{tsamanta@iiap.res.in}}\\ 
$^{2}$ School of Earth and Space Sciences, Peking University, China; {\color{blue}{huitian@pku.edu.cn}} \\
$^{3}$ Center of Excellence in Space Sciences, IISER Kolkata, India; {\color{blue}{dipu@iiap.res.in}}\\ 
% $^{4}$ National Astronomical Observatory of Japan, Tokyo, Japan; {\color{blue}{yukio.katsukawa@nao.ac.jp}}\\
$^{4}$ Harvard-Smithsonian Center for Astrophysics, Cambridge, USA; {\color{blue}{nschanche@cfa.harvard.edu }}}

%%%---------------------Abstract-------------------------%%%%
\begin{abstract}
Recent high-resolution observations reveal that subarcsecond bright dots~(BDs) with sub-minute lifetimes appears ubiquitously in 
the transition region~(TR) above sunspot penumbra. The presence of penumbral micro-jets~(PMJs) in the 
chromosphere have also been reported earlier. 
It was proposed that both the PMJs and BDs are formed due to magnetic reconnection process and may play an important role in heating of the penumbra.
Using simultaneous observation of the chromosphere from the Solar Optical Telescope~(SOT) aboard Hinode and the TR from the 
Interface Region Imaging Spectrograph~(IRIS), 
% (IRIS) 
we study the dynamics of BDs and their relation with PMJs.  
% Using  combined observations, we compare these two features and try to identify if they have a common origin. 
We find two types of BDs, one which is related to PMJs and 
the others which do not show any visible dynamics in the SOT~\ion{Ca}{2}~H  images. From a statistical analysis we show that these two 
types have different properties. 
The BDs which are related to PMJs always appear at the top of the PMJs, the vast majority of which show inward motion and 
originate before the generation of the PMJs. 
These results may indicate that the reconnection occurs at the lower coronal/TR height and initiates PMJs at the chromosphere.
This formation mechanism is in contrast with the currently believed formation of PMJs by reconnection in the~(upper) 
photosphere between differently inclined fields.
% These behavior is puzzling and can not be explain by a simple reconnection model. 
% In this work we explore the relationship and propose a new picture for formation of BDs and their relation to PMJs.
\end{abstract}

% \online{: animation, color figures}
%%%%%-----------------Introductory section-----------------%%%%%

\section{Introduction}
Sunspots are regions of concentrated strong magnetic fields comprising of dark central region, umbra surrounded by a less darker region called the penumbra. 
% The magnetic fields in the umbra are more vertical and in the penumbral region, where they are more inclined. 
% Though the sunspots have been observed over centuries, their fine structures and their counterparts in the higher solar atmosphere 
% have been revealed in recent decades with the advancement of high-resolution and multi-wavelength observations.
% With the advancement of high-resolution instruments, fine structures and their dynamics are being studied. 
Penumbral microjets (PMJs) are one of the prominent fine-structure dynamical features observed in the sunspot at the chromospheric height.
Using the SOT \citep{2008SoPh..249..167T} \ion{Ca}{2}~H  filter images, % aboard Hinode, %\citep{2007SoPh..243....3K}, 
\citet{2007Sci...318.1594K} first reported that these micro-jets
occur ubiquitously above  penumbra. %In \ion{Ca}{2}~H  filter images and also in time-difference movies the presence of these jets are clearly visible. 
They have lengths of 1~to~4~Mm
% , widths of a few hundred kilometers 
and lifetimes of up to a minute. They generally move very fast and have apparent speeds over 100~km~s$^{-1}$.
The magnetic field in the chromospheric penumbral region consists of a combination of spines (more
vertical field) and interspines (more horizontal field) \citep{1993ApJ...418..928L,2004A&A...427..319B,2011Sci...333..316S,2012A&A...540A..19S,2013A&A...557A...5H,2013A&A...557A..25T,2015A&A...583A.119T}.
\citet{2007Sci...318.1594K} found that PMJs generally originate near the bright structures in
between two dark penumbral filaments and propagate upward along the direction of the spine.
So, they proposed that PMJs could originate as a result of magnetic reconnection between the spine and interspine magnetic fields. 

%%%%% Figure Observation
%%%%%%%%%%%%%%%%%%%%%%%%%%%%%%%%%%%%%%%%%%%%%%%%%%%%%%%%%%%%%%%%
\begin{figure*}[!htbp]
\centering
\includegraphics[angle=90,clip,width=18.0cm]{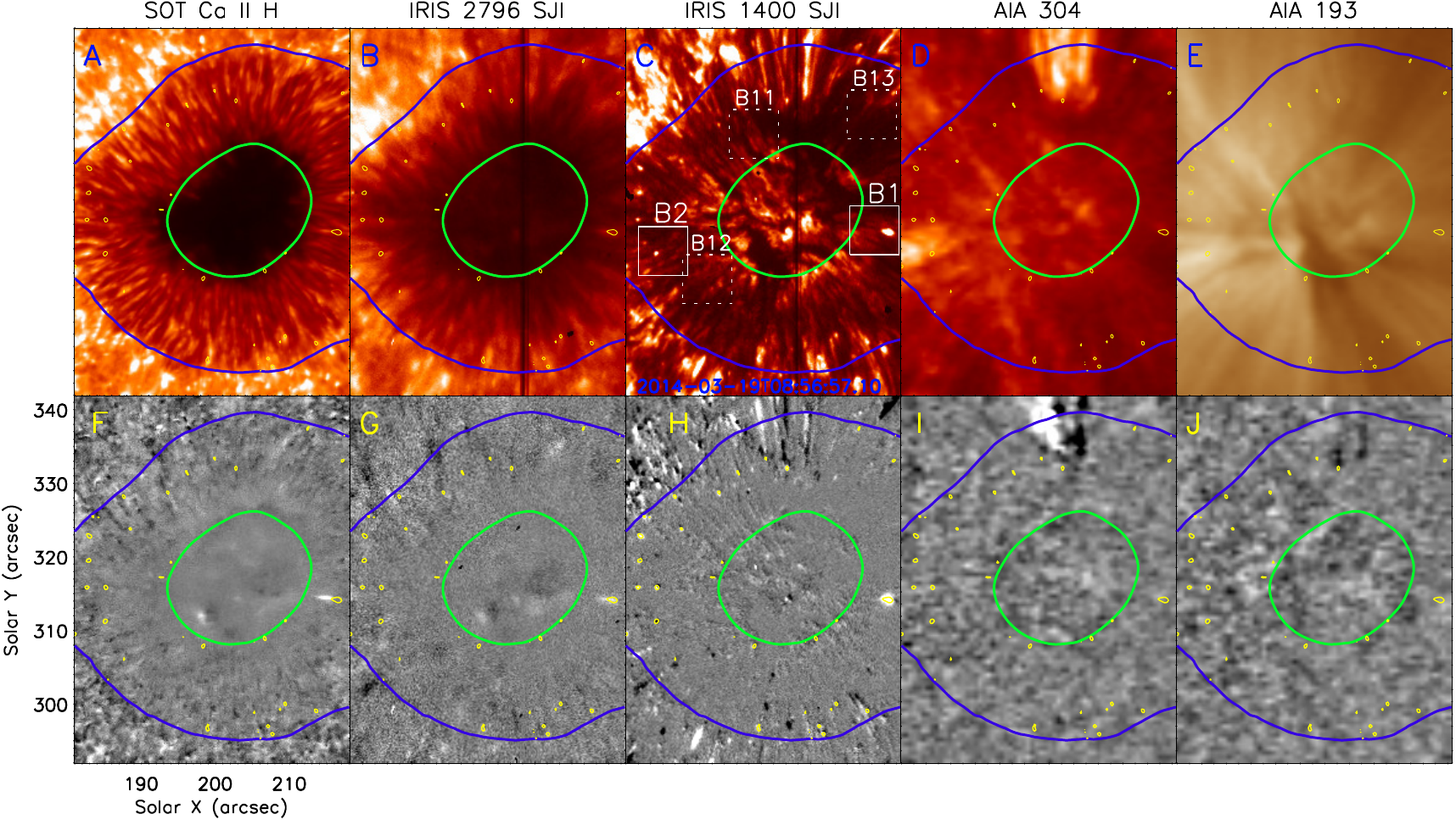}
\caption{A-E: Image of a sunspot as seen by the SOT, IRIS and AIA around 08:57 UT on 19 March 2014. F-J: Show the running difference images. 
The green and blue contours, derived form a time-averaged
smoothed SOT image, represent the umbra and penumbra of the sunspot, respectively. 
The yellow contours mark the locations of bright dots (BDs).
Evolution of the BD inside the box B1 and B2 (as marked in the 1400~\r{A}) is shown in the Figure~\ref{ev_b1ab2} (animated figures corresponding to B1, B2, B11-B13 are available online.)
% The BD inside the B1 box is related to a PMJ. The BD inside box B2 is not related to any visible dynamics in the SOT.
An animation of this full figure is also available.
}
\label{sot_iris_aia} 
\end{figure*}
%%%%%%%%%%%%%%%%%%%%%%%%%%%%%%%%%%%%%%%%%%%%%%%%%%%%%%%%%%%%%%%%
%
%The Interface Region Imaging Spectrograph (IRIS) \citep{2014SoPh..289.2733D} provides unique high-resolution data
%of the highly dynamic, less understood, chromosphere and transition region (TR). 
Using the IRIS \citep{2014SoPh..289.2733D} observations, \citet{2014ApJ...790L..29T} found the presence of
subarcsecond Bright Dots (BDs) above sunspot penumbra in the TR. 
They appear ubiquitously and mostly have a lifetime less than few minutes. 
They sometimes appear slightly elongated
along the penumbral filaments and also move along the filaments with speeds of 10--40~km~s$^{-1}$. 
They proposed that some of them could be due to impulsive reconnection in the TR and chromosphere at footpoints of coronal magnetic loops and others
are probably due to falling plasma.
It is still unclear how BDs are formed and if they show any signatures in the lower and upper atmosphere.

% Using 
% % multi-wavelength imaging observations from 
% the Swedish 1-m Solar Telescope and IRIS observations, 
\citet{2015ApJ...811L..33V} studied the multi-wavelength signatures of PMJs and found that PMJs show spatial offset from the chromosphere to
TR in the direction of the PMJ.  
Hence, they proposed that PMJs may progressively heat up to TR temperature. 
\citet{2016ApJ...816...92T} could not find noticeable signature of normal PMJs in any Atmospheric Imaging Assembly (AIA) \citep{2012SoPh..275...17L} passbands,
although a few strong and large PMJs show BD/BD-like signatures in the  1600~\AA\ and High-resolution Coronal imager (Hi-C) 193~\AA\ images.  
% They proposed that larger PMJs originate from magnetic reconnection between the
% opposite-polarity field of spines and filament tails and produces significant heating in the TR and corona.
\citet{2016ApJ...822...35A} found that BDs observed in Hi-C are on average slower, dimmer, larger in size and longer lived than
IRIS penumbral BDs. They also found that most of the BDs observed in Hi-C 193~\AA\ may correspond to TR. 
\citet{2016ApJ...829..103D} found that the locations of most of the penumbral BDs show downflow. % and originate at the locations of the darker fibrils in the chromosphere. 
Their statistical analysis shows that BDs do not have consistent brightening response in the chromosphere. 
Following \citet{2014ApJ...790L..29T}, they also suggested that TR penumbral BDs are manifestation of falling plasma from coronal heights along more vertical and dense magnetic
loops or small-scale impulsive magnetic reconnection at TR or higher heights.
% They could be also produced by small-scale impulsive magnetic reconnection at TR or higher heights. 
\citet{2014ApJ...789L..42K} and \citet{2016A&A...587A..20C} observed heating events which are associated with the strong downflows in the TR and proposed a similar scenario.
\citet{2016ApJ...823...60B} have studied  small transient brightening events in the penumbra and found that it show redshifts  in the \ion{Si}{4}~1402.77~\AA\ line, an inward motion towards the umbra in IRIS 1400~\AA\ images and have a multi-thermal component. They proposed the triggering mechanism as magnetic reconnection at low coronal heights.
A component of the plasma from the reconnection site may move downward and reach the TR, which is confirmed 
by the observed redshifts in the \ion{Si}{4}~1402.77~\AA\ line and the inward motions as seen in the IRIS 
1400~\AA\ images. Finally, it reaches  the chromosphere and appears as ribbon-like brightening. 
% \citet{2006SoPh..234...41K}, \citet{2012ApJ...751..152J} and \citet{2013ApJ...771...21W} proposed that  
% this kind of small magnetic reconnection occurs  in the low corona. 

In this work, while combining multi-wavelength observations covering chromosphere, TR/corona  we try to find the source of the BDs and their relation to PMJs. 
%%%%%%%%%%%%%%%%%%%%%%%%%%%%%%%%%%%%%%%%%%%%%%%%%%%%%%%%%%%%%%%%
\begin{figure*}
\centering
\includegraphics[angle=90,clip,width=9.0cm]{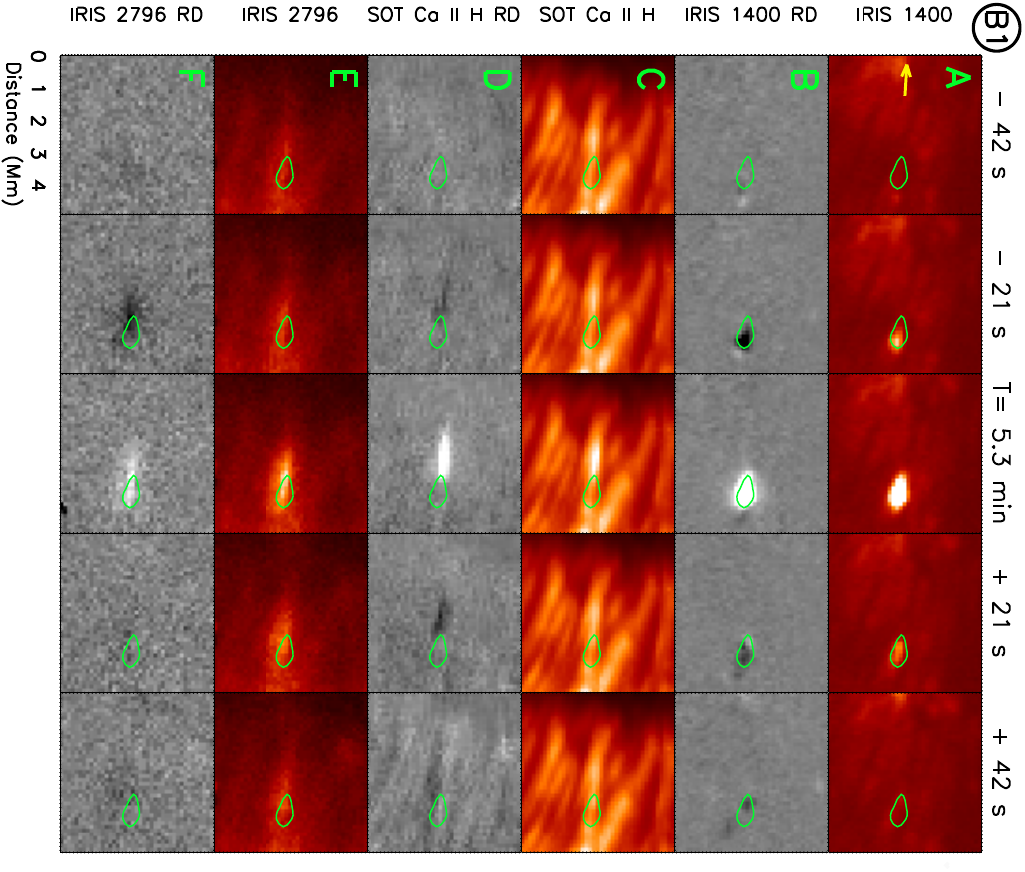}\includegraphics[angle=90,clip,width=9.0cm]{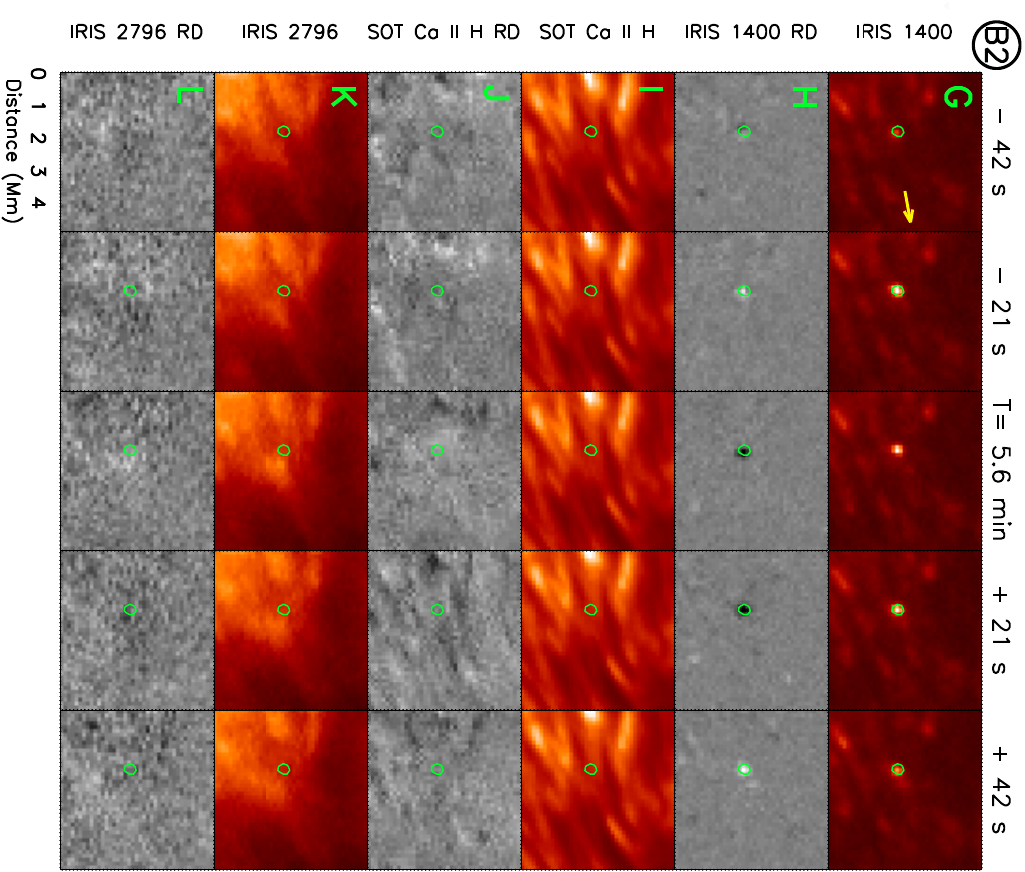}
\caption{\textcircled{\textbf{\tiny{B1}}} show the temporal evolution of a BD and a PMJ inside the B1 region (as marked in the Figure~\ref{sot_iris_aia}) as seen in different filtergram images and their running difference images.   
The green contour is derived from the top middle panels to show the location of the BD. Yellow arrow indicates the direction of the sunspot center. 
Similarly, \textcircled{\textbf{\tiny{B2}}} show the temporal evolution of the B2 region (as marked in the Figure~\ref{sot_iris_aia})}
\label{ev_b1ab2} 
\end{figure*}
%%%%%%%%%%%%%%%%%%%%%%%%%%%%%%%%%%%%%%%%%%%%%%%%%%%%%%%%%%%%%%%%
%%%%%--------------------Main body-------------------------%%%%% 
% ----------------------------1---------------------------------
\section{Data analysis and Results}
\subsection{Observation and Data Reduction}
We use the data obtained from a coordinated observation taken on 19 March 2014, using the HINODE/SOT (HOP-250) \citep{2008SoPh..249..167T}, IRIS (IRIS-3840007453) \citep{2014SoPh..289.2733D} 
and the AIA \citep{2012SoPh..275...17L} on board the Solar Dynamics Observatory. 
The coordinated observation between all the instruments was performed from 08:51~UT to 09:46~UT. 
We selected a FOV of $37''$x$50''$ centered at $200''$ and $317''$. The FOV is limited due to the SOT observations.
IRIS Slit-jaw images (SJI) centered at 2796~\r{A} and 1400~\r{A}  are dominated by the \ion{Mg}{2}~k and \ion{Si}{4} emission lines, respectively.
The SOT filter is dominated by \ion{Ca}{2}~H  line, which forms at the lower chromosphere where the temperature is below 10$^{4}$ K. 
The IRIS 2796~\r{A} images are dominated by the emission from a plasma at temperatures $\sim$10,000-15,000 K and represent the middle chromosphere.
The IRIS 1400~\r{A} passband is sensitive to TR $\sim$60,000-80,000 K.
AIA filtergram images centered at 304~\r{A} (sensitive to $\sim$0.05~MK) and 193~\r{A} ($\sim$1.25~MK) are dominated 
by \ion{He}{2} and \ion{Fe}{12} emission lines, respectively. 
% The AIA 304~\r{A} and AIA 193~\r{A} filter response functions peak at 0.05~MK and 1.25~MK, respectively.
%We used IRIS Level~2 data available online (OBSID 3840007453). 
% which has dark current, flat field and geometrical corrections etc. 
SOT \ion{Ca}{2}~H  data was taken with 0.3~sec exposure and 1.58~sec cadence.
IRIS SJI's were obtained with 4~sec exposure and a cadence of 10.5~sec.
AIA 304~\r{A} and 193~\r{A} observations were taken with 2 sec exposure and 12 sec cadence. 
% All the AIA passbands were prepped and normalized with the aia prep routine in SolarSoft. 
AIA data were then co-aligned and de-rotated to the start time (08:51~UT) of observation to compensate for the solar rotation.
The pixel size of SOT, SJI's and AIA are $0.109''$, $0.166''$ and $0.6''$, respectively.  
We have interpolated SOT and AIA data in space and time to match the IRIS SJI cadence (10.50 sec) and spatial resolution ($0.33''$ ).
The IRIS SJIs and SOT images were co-aligned using IRIS 2796~\r{A} and SOT \ion{Ca}{2}~H  images.
The time difference between the SOT and IRIS image is of the order of sub-seconds.
After that IRIS and AIA were co-aligned using IRIS 1400~\r{A} and AIA 1600~\r{A} images \citep{2015ApJ...806..172S}.
% We should point out that aligning images form different spacecrafts is a difficult task as
% there is no common filter and we may haven't achieve sub-pixel accuracy. 

Figure~\ref{sot_iris_aia} shows the image of a sunspot, where the bottom rows (G-K) correspond to running difference images. 
% The green and blue contours, derived form a time-averaged smoothed SOT image, represent the umbra and penumbra of the sunspot, respectively. 
The yellow contours show the location of BDs in the sunspot penumbra. 
These contours are obtained from the 1400~\r{A} image after subtracting the same image with a 8x8 pixel smoothing.
Evolution of two BDs inside the box B1 and B2 (as marked in 1400~\r{A}) are shown in the Figure~\ref{ev_b1ab2} (also in online movie). 
The BDs inside the box B1 including box B11, B12 and B13 are related to PMJs.
The BD inside B2 is not related to any visible dynamics in the SOT. 
%%%%%%%%%%%%%%%%%%%%%%%%%%%%%%%%%%%%%%%%%%%%%%%%%%%%%%%%%%%%%%%%
\subsection{Temporal evolution of BDs and PMJs}
%%%%%%%%%%%%%%%%%%%%%%%%%%%%%%%%%%%%%%%%%%%%%%%%%%%%%%%%%%%%%%%%
\begin{figure*}
\centering
\includegraphics[angle=90,trim = 0mm -5mm 0mm 0mm,clip,width=9.0cm]{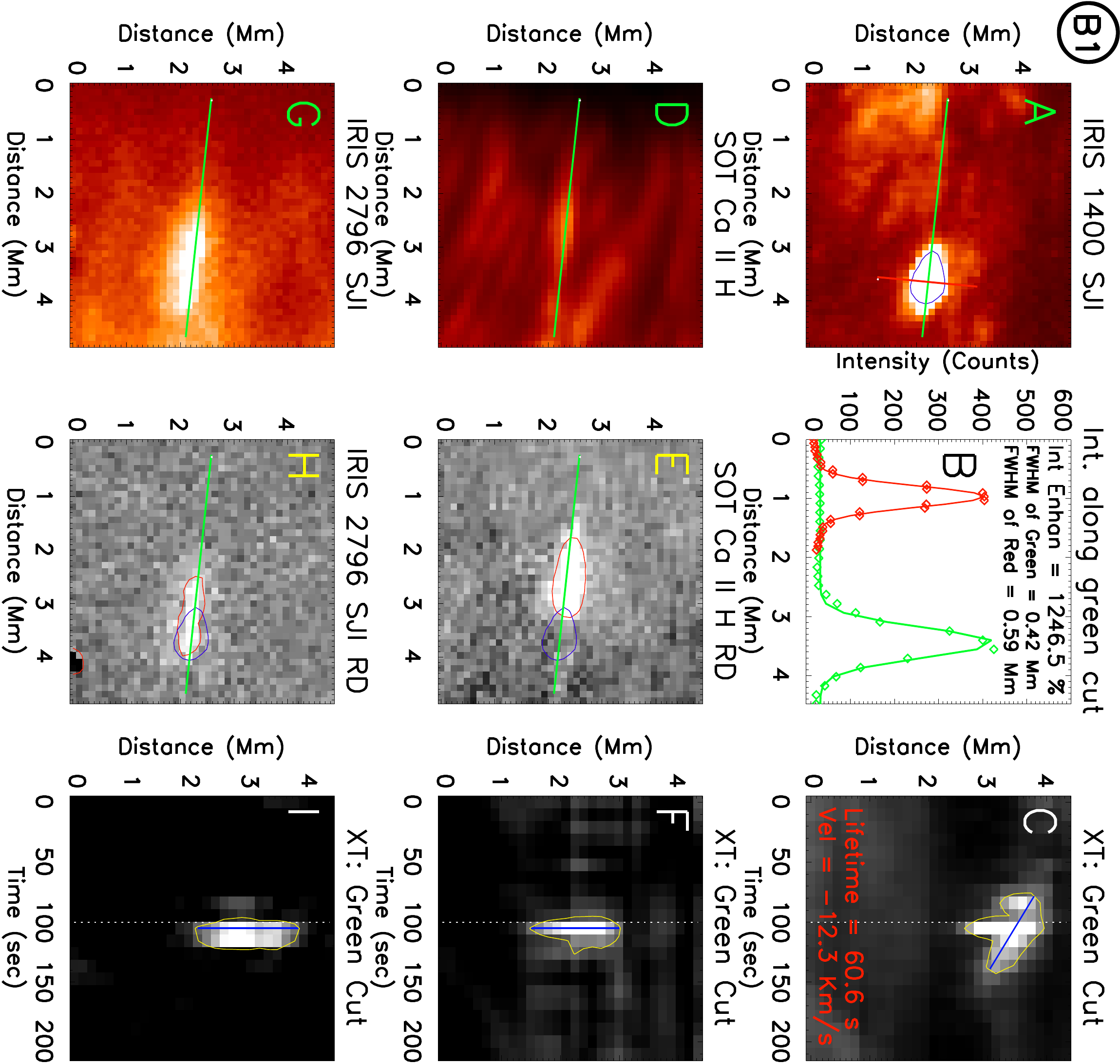}\includegraphics[angle=90,trim = 0mm 0mm 0mm -5mm,clip,width=9.0cm]{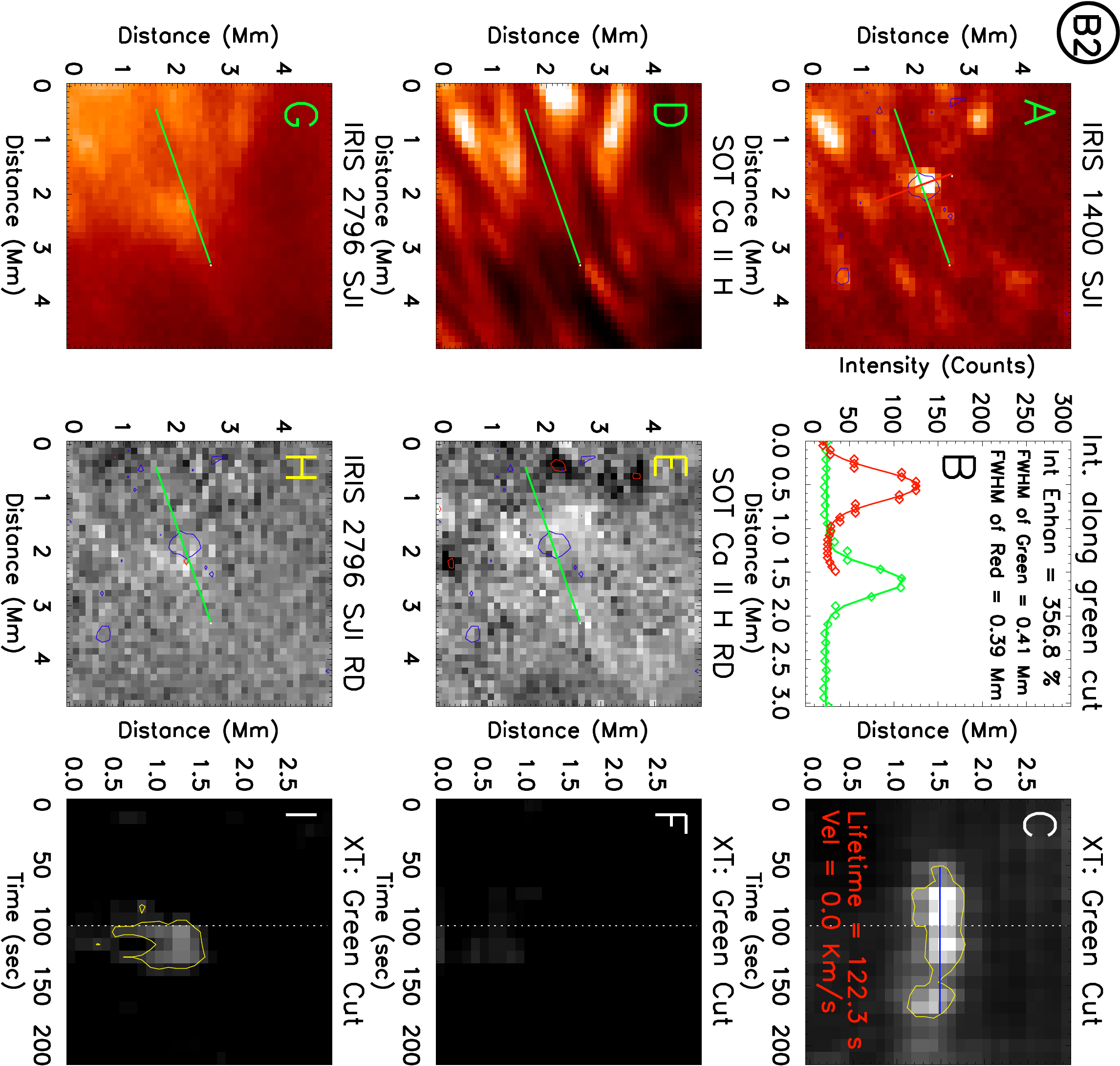}
\caption{\textcircled{\textbf{\tiny{B1}}} show the procedure of determining the physical parameters of the BD and the PMJ inside the B1 region (as marked in the Figure~\ref{sot_iris_aia}). 
(A) show the BD enclosed by the blue contour as seen in the 1400~\r{A}. The green slit and red slit is placed along the radially outward direction and its perpendicular direction, respectively.
(B) show the intensity profile along the green and red slit with diamond symbol. The solid line is a Gaussian fit to the profiles to determine the width and intensity enhancement of the BD.
(C) show the temporal evolution (XT plot) along the green cut. Yellow contour show the region above 2$\sigma$ intensity enhancement. The slop of the blue line is used to determine the 
plane of the sky velocity of the BD.
(D) show \ion{Ca}{2}~H  image. The green line  is the slit to determine XT plot. 
(E) show the \ion{Ca}{2}~H  image after subtracting the previous frame. Red contour show the location of the PMJ. 
(F) show the XT plot for the green cut of \ion{Ca}{2}~H images.
(G) show 2796~\r{A} image. (H)-(I) are similar to (E)-(F), respectively, but for the 2796~\r{A} channel.
% The green line is the slit to determine XT plot. 
% (H) show the 2796 image after subtracting from the previous frame. Red contour show the location of the PMJ. 
% (I) show the XT plot for the green cut of IRIS 2796 images.
Similarly, \textcircled{\textbf{\tiny{B2}}} show for the BD inside B2 region (as marked in the Figure~\ref{sot_iris_aia}).}
\label{xt_b1ab2} 
\end{figure*}
%%%%%%%%%%%%%%%%%%%%%%%%%%%%%%%%%%%%%%%%%%%%%%%%%%%%%%%%%%%%%%%
Figure~\ref{ev_b1ab2} depicts the temporal evolution of two BDs.
\textcircled{\textbf{\tiny{B1}}}~shows the evolution of the BD inside the B1 box in the Figure~\ref{sot_iris_aia}.
Panel A, C and E show the evolution as seen in the IRIS 1400~\r{A},  SOT \ion{Ca}{2}~H  and IRIS 2796~\r{A}, whereas the B, D and F correspond to running difference images. 
The clear presence of a BD is seen in the in the 1400~\r{A}. A jet like feature is also seen in the \ion{Ca}{2}~H  and 2796~\r{A} images. 
Similarly, \textcircled{\textbf{\tiny{B2}}} show the evolution of the B2 region as marked in the Figure~\ref{sot_iris_aia}. Here, we can clearly find the presence of a BD in the 1400~\r{A} channel 
but no clear signature of intensity enhancement  in the \ion{Ca}{2}~H.
We have identified several BDs and performed a statistical analysis to compile the properties of these BDs and their signatures in the other channels.
At first, we identify manually an isolated BD and look for the time frame where it shows maximum intensity enhancement. 
We use that as our central frame and make a movie (see animated figure of Fig.~\ref{ev_b1ab2}, covering a 5x5~Mm region for a duration of 210 sec) to follow the  BDs and their direction of  propagation. 
The BDs often show apparent movements along the bright filamentary structures roughly in the radial direction of the sunspot. 
% Though some of the BDs do not follow the exact radial direction.
%%%%%%%%%%%%%%%%%%%%%%%%%%%%%%%%%%%%%%%%%%%%%%%%%%%%%%%%%%%%%%%%
\begin{figure*}[!htbp]
\centering
\includegraphics[angle=90,clip,width=15cm]{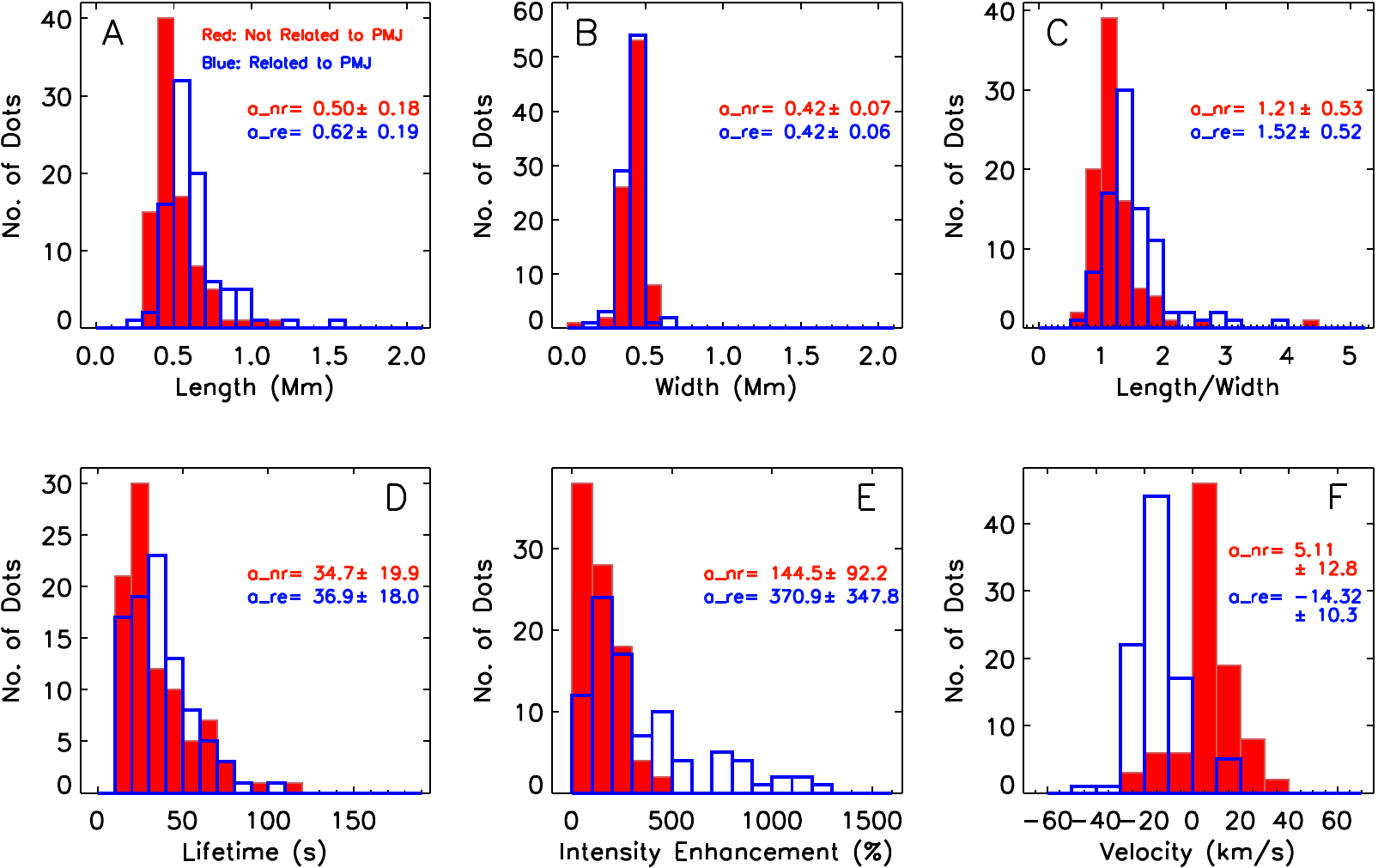}
\caption{(A-F) show the distribution of different physical parameters of BDs.
Red represents the BDs which do not have any connection to the PMJs where as blue represents those which are found on the top of the PMJs.
(A) show the of the length (FWHM of the intensity profiles along the green slit), 
(B) the width (FWHM of the intensity profiles along the red slit), 
(C) the ratio of length and width, 
(D) the lifetime,
(E) the intensity enhancement
and (F) the plane of the sky speed of the BDs.
The average value of each parameters and also their standard deviation are printed (a$\_nr$ and $a\_re$). 
}
\label{stat_bd} 
\end{figure*}
%%%%%%%%%%%%%%%%%%%%%%%%%%%%%%%%%%%%%%%%%%%%%%%%%%%%%%%%%%%%%%%%

Now to find the properties of the BDs, we follow a similar method as used by \citet{2014ApJ...790L..29T}. 
Two examples of the analysis are shown in Figure~\ref{xt_b1ab2}~\textcircled{\textbf{\tiny{B1}}}~\&~\textcircled{\textbf{\tiny{B1}}}. 
We have studied 180 penumbral BDs and try to find if they have any signature in the SOT \ion{Ca}{2}~H and IRIS 2796~\r{A} channels.
We use the central image to compute the intensity enhancement, length and width of a BD, as shown in Figure~\ref{xt_b1ab2}~\textcircled{\textbf{\tiny{B1}}}. We plot the intensities along the green cut and red cut (panel A). The  green cut is placed  along the
radial direction and the red cut is  perpendicular to it. 
The intensity values are shown by red  and green diamond symbols. A fitted Gaussian is also overplotted with the same color. 
The FWHM of the green profile provides an estimate of the length of the BD, whereas  the red profile provides the width. 
Percentage intensity enhancement is calculated from the peak intensity and the linear background of the fitted Gaussian. 
The smallest between the two values (green and red) measures the intensity enhancement. 
We plot space-time (XT) diagram (panel~C) to study their temporal behavior and to measure their apparent speed.
 A 2$\sigma$ intensity contour is drawn on the XT map. The inclined blue line is drawn inside
the contour to measure the apparent velocity and lifetime of the BD. The starting time of the blue line is considered
as the initiation of the BD. The long intensity strip at the center of the XT map is due to high intensity enhancement at that particular time frame.
We follow a similar method for the \ion{Ca}{2}~H and 2796~\r{A} channels (D and G). 
% E and H show the running difference images. 
The intensity enhancement in these channels is only a few percent and very difficult to find the location
of the intensity enhancement. So, we use the running difference images to place our green and red cut (see E and H) on the original images.
%%%%%%%%%%%%%%%%%%%%%%%%%%%%%%%%%%%%%%%%%%%%%%%%%%%%%%%%%%%%%%%%
\begin{figure*}
\centering
\includegraphics[angle=90,clip,width=16cm]{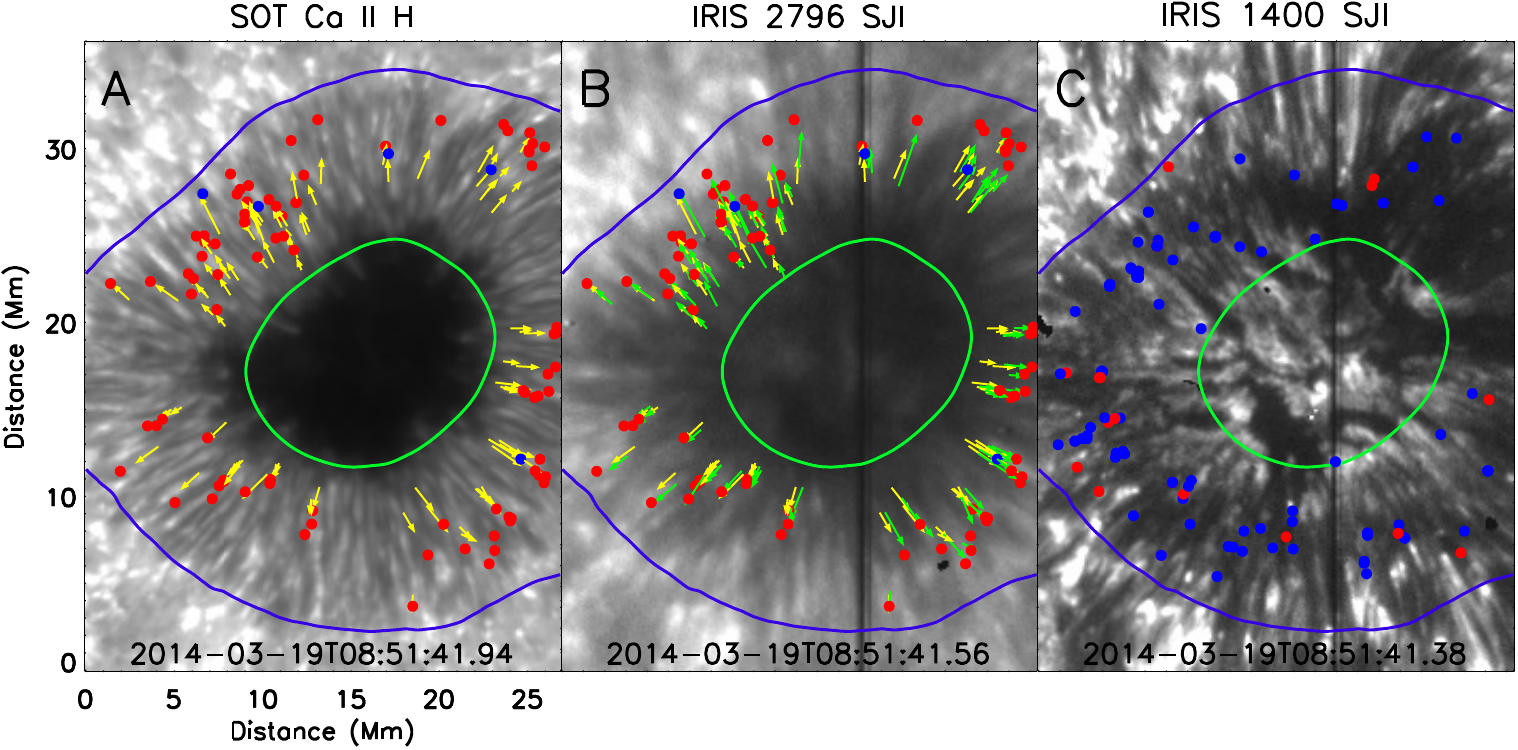}
\includegraphics[angle=90,clip,width=15cm]{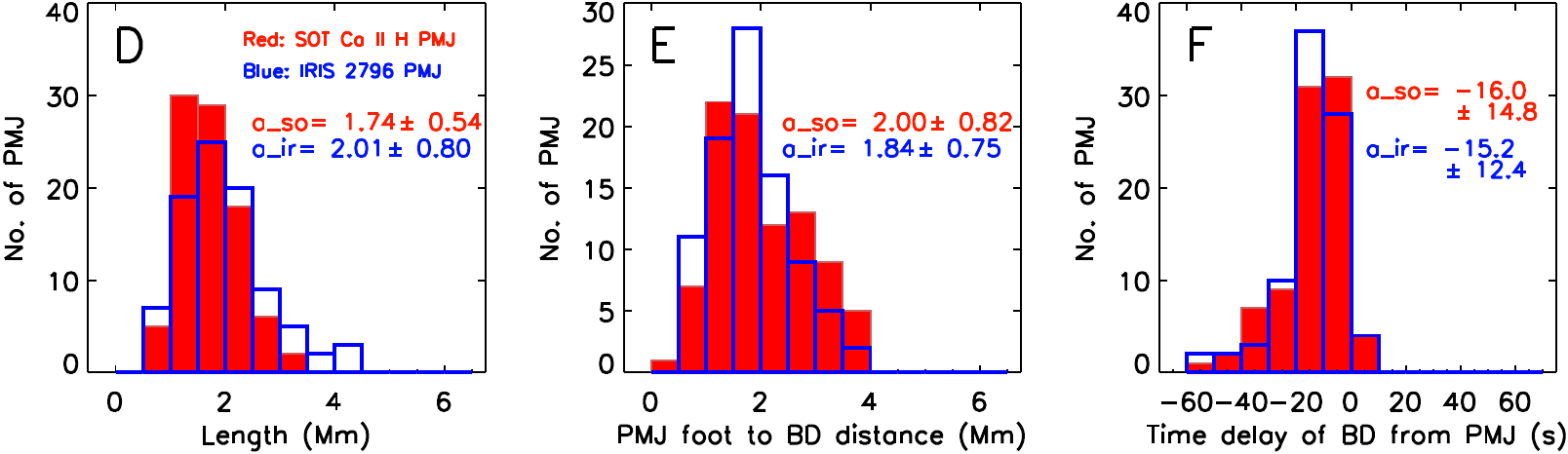}
\caption{(A): The yellow arrows show the location of the PMJs as found in the \ion{Ca}{2}~H images and the dots (red/blue) shows the location of 
the BDs as seen in the 1400~\r{A} channels over the \ion{Ca}{2}~H image.
% on top of SOT \ion{Ca}{2}~H  image. 
Red dots represent the BDs which moves inward direction (opposite of the PMJ's direction) and blue represent the BDs which moves outward (along the direction of PMJs).
(B): The yellow arrow and the dots (red/blue) show the location of \ion{Ca}{2}~H PMJs and 1400~\r{A} BDs over the 2796~\r{A} SJI. 
The green arrows represent the PMJs as observed in the 2796~\r{A} images.
(C): The BDs which do not have any visible counterparts in the \ion{Ca}{2}~H images are shown over the 1400~\r{A} image.
(D) show the distribution of the length of the PMJs.
(E) show the distribution of the distance between the footpoint of the PMJs and starting location of the BDs.
(F) show the distribution of the time delay of the origination of BD after PMJ.
A negative value means that a BD originates earlier than a PMJ.
All the PMJs and BDs are observed over the total observing window whereas the background images are taken at t=0.
The mean values of each parameter and also their standard deviation are printed. The $a\_so$ and $a\_ir$ represents the mean value as obtained from the SOT \ion{Ca}{2}~H and IRIS 2796~\r{A}, respectively. 
}
\label{pos} 
\end{figure*}
% %%%%%%%%%%%%%%%%%%%%%%%%%%%%%%%%%%%%%%%%%%%%%%%%%%%%%%%%%%%%%%%%
%
The XT map is produced after removing a smoothed background from each images. An elongated bright structure is seen in the XT plot. These have been conventionally referred as penumbral micro-jets (PMJs) by \citet{2007Sci...318.1594K}. The lifetimes of these jets are very small.
Most of them appear as sudden brightening in the XT plot and hence it is very difficult to determine their direction of propagation and the speed.
% As we use a cadence of 10.5 sec, it is difficult to calculate their speed. 
In this work we do not focus on the speed of these jets.
Using earlier convention we have assumed that these brightenings are propagating outward form the sunspot, though
it is not  clearly established from our analysis that the bright structures are moving outward from the sunspot.
% it is very difficult to conclude from our analysis.
% A long jets like feature is clearly seen in the XT plot.
We use a 2$\sigma$ intensity contour to find the location and extent of the PMJ. The blue line is drawn along the direction of intensity enhancement.
The starting time of the blue line is considered as the initiation and the length of the line is considered as the length of the
PMJ. We can clearly see that the BD appears before the PMJ in this example.
% The red cut is used to measure the width of the jets.

Similarly, in Figure~\ref{xt_b1ab2}\textcircled{\textbf{\tiny{B2}}}, we  study another BD.
% The results are also shown in the Figure~\ref{xt_b1ab2}\textcircled{\textbf{\tiny{B1}}}.
Following same methodology we study the evolution of the BD.
% and try to find if they have any counterparts in the \ion{Ca}{2}~H and 2796 channels.
We could not find a clear signature of the BD in the \ion{Ca}{2}~H though a faint signature is seen in the 2796~\r{A} channel.
% In our study, 88 dots shows clear signatures in the lower channels but 96 dots does not show any distinguishable signatures in the SOT.
% These two types of BDs shows different properties. 
In the following subsection a statistical study is performed  to find out if there are two types BDs present in our data.
%%%%%%%%%%%%%%%%%%%%%%%%%%%%%%%%%%%%%%%%%%%%%%%%%%%%%%%%%%%%%%%%
\subsection{Statistical behavior of the BDs}
We analyze the properties of many BDs and their signature in the other channels.
% We have identified  90 BDs with a clear signature in the \textbf{2796~\r{A}  and \ion{Ca}{2}~H} whereas many do not show any distinguishable signatures in the \ion{Ca}{2}~H images. 
Out of the total 180 events, 90 are identified as having a correlated signature in the Ca II H and 2796~\r{A} channels, while the other 90 do not.
% Out of that we present statistical properties of this sample of 90 BDs which do not show signatures in \ion{Ca}{2}~H images. 
We have separated these two classes so as to find out if they have any significant differences in their physical properties. 
% and if we could comment about their origination.
Figure~\ref{stat_bd} (A)-(F) show the distributions of the observed parameters. 
The red color in the histograms represent the BDs which do not have a counterpart in the \ion{Ca}{2}~H images whereas the 
blue represent the BDs which are related to the \ion{Ca}{2}~H PMJs.
The mean value of each of the parameters for both the types are also printed.
We find that the BDs which are related to PMJs, generally, are longer and have higher intensity enhancements.  
They are also more elongated (the ratio between length and width is higher).
One of the distinguishable feature  is  that most of the BDs which are related to PMJs show negative velocities (drifts towards the center of the sunspot).

\subsection{Statistical behavior of PMJs and their connection to BDs}
We have also calculated a few physical parameters of PMJs. 
% We obtained the parameters of the PMJs as seen in both the 2796 and \ion{Ca}{2}~H channels to compare if they show 
% any differences in their behavior. 
The position and length of each PMJs are measured in both the 2796~\r{A} and \ion{Ca}{2}~H channels.
The starting point of the PMJs is considered as the location close to the center of the sunspot.
The locations of the PMJ as seen in the \ion{Ca}{2}~H and 2796~\AA are marked (yellow arrows) in the Figures~\ref{pos}A, B respectively.   
% The arrow indicates the direction of the propagation of the PMJs and 
The length of the arrow measures the length of the PMJs. 
The BDs which are related to these PMJs are also shown by the small dots. Red represents the dots which
moves inward and blue represents those that move outward from the sunspot center.
% Figure~\ref{pos}B show the image of IRIS 1400 and overplotted yellow arrows are PMJs as seen in the 
% SOT and green arrows represent the PMJs as seen in the IRIS 2796 images.
The PMJs as seen in 2796~\r{A} generally show spatial offset from the \ion{Ca}{2}~H PMJs along the PMJs direction (outer penumbral side).
In Figure~\ref{pos}C, we show the location of the BDs which are not related to any PMJs.
Figure~\ref{pos}(D) show the distribution of the length of the PMJs as seen in the 2796~\AA\ and \ion{Ca}{2}~H images. The PMJs in \ion{Ca}{2}~H have similar value as reported by \citet{2007Sci...318.1594K} though they appear little longer in the 2796~\AA\ images.
Figure~\ref{pos}(E) show the distribution of the length between the BDs and the footpoint of the PMJs and
Figure~\ref{pos}(F) show the time delay between the appearance of the BDs and PMJs. A negative value means BDs appear earlier than the PMJs. We find that most of the BDs appear before the PMJs.

\section{Summary}
We have analyzed the properties of 180 BDs as seen in the 1400~\r{A} images above a sunspot. 
% They generally appear slightly elongated along the bright filamentary structures of the penumbra and show an apparent movement both inward and outward direction from the sunspot. 
Using our coordinated observation, we have also positively identified PMJs in the SOT \ion{Ca}{2}~H and IRIS 2796~\AA\ ~images.
We find that 90 BDs are related to PMJs 
% and may originate from a common process 
whereas others are not.
A detail analysis show that the BDs which are related are generally longer and more elongated and have higher intensity enhancements.  
Most of these BDs show negative velocities (moves inward direction) and appear at top of the PMJs.
These BDs are found to be appearing before the generation of the PMJs.
% The results obtained from our statistical analysis could be explained by the proposed  method of \citet{2016ApJ...823...60B}. 
These results may indicate that BDs could originate from a magnetic reconnection occurring at low coronal heights and/or due to falling plasma.
A component of the plasma from the reconnection site may move downwards and reach the TR and show inward motion as seen in the IRIS 1400~\AA\ images. Finally, it reaches 
the chromosphere and appears as ribbon-like brightening (PMJ) which could explain the time delay in the appearance of the PMJs and their short lifetime. \citet{2006SoPh..234...41K}, \citet{2012ApJ...751..152J} and \citet{2013ApJ...771...21W} proposed that  
 this kind of small magnetic reconnection occurs  in the low corona. This is in contrast to the reconnection model as proposed by \citet{2007Sci...318.1594K}.
% We should point out that the 
The progressive heating mechanism of the PMJs as proposed by \citet{2015ApJ...811L..33V} also 
may not explain the inward motions of the BDs, appearance of BDs before PMJs and the longer lifetime of the BDs.
It is still unclear how the non-related BDs originate. 
 
%%--------------------------------------------------------------------------------------------------------------
\acknowledgments
We thank Yukio Katsukawa for useful discussions and suggestions. H. Tian is supported by the Recruitment Program of Global Experts of China, NSFC under grant 41574166, and the Max Planck Partner Group program. 
H.Tian thanks ISSI Bern for the support to the team ``Solar UV bursts: a new insight to magnetic reconnection''. We thank the Hinode, IRIS and SDO team for proving the data in the public domain.
Hinode is a Japanese mission developed and launched by ISAS/JAXA, with NAOJ as domestic partner and NASA and STFC (UK) as international partners. It is operated by these agencies in co-operation with ESA and NSC (Norway).
IRIS is a NASA small explorer mission developed and operated by 
LMSAL with mission operations executed at NASA Ames Research center 
and major contributions to downlink communications funded by the Norwegian Space Center (NSC, Norway) through an ESA PRODEX contract.

% 
% \bibliographystyle{apj}
% \bibliography{references}

\end{document}